# Large phonon-drag thermopower polarity reversal in Ba-doped KTaO$_3$

Mohamed Nawwar,[1] Samuel Poage,[1] Tobias Schwaigert,[2] Maria N. Gastiasoro,[3] Salva Salmani-Rezaie,[2] Darrell G. Schlom,[2,4,5] Kaveh Ahadi,[1,6] Brandi L. Wooten,[1,†*] Joseph P. Heremans[1,7,8, *‡]

1. Department of Materials Science and Engineering, The Ohio State University, Columbus OH 43210, USA
2. Department of Materials Science and Engineering, Cornell University, Ithaca, NY 14853, USA
3. Donostia International Physics Center, 20018 Donostia-San Sebastian, Spain
4. Kavli Institute at Cornell for Nanoscale Science, Ithaca, NY 14853, USA
5. Leibniz-Institut fur Kristallzüchtung, 12489 Berlin, Germany
6. Department of Electrical and Computer Engineering, The Ohio State University, Columbus, OH 43210, USA
7. Department of Mechanical and Aerospace Engineering, The Ohio State University, Columbus OH 43210, USA
8. Department of Physics, The Ohio State University, Columbus OH 43210, USA

\* Corresponding author
† Contact author: Brandi.L.Wooten.civ@Army.mil
‡ Contact author: Heremans.1@osu.edu

**ABSTRACT**: This study reports the observation of phonon-drag thermopower polarity reversal in Ba-doped KTaO$_3$ thin films, mediated by electron-phonon Umklapp scattering. Epitaxial films with distinct carrier concentrations ($3.7\times10^{20}$ cm$^{-3}$ and $4.9\times10^{19}$ cm$^{-3}$) were grown via molecular-beam epitaxy. In heavily doped samples, where the Fermi surface spans 80% of the Brillouin zone, the Umklapp condition is satisfied, reversing electron momentum. This manifests as a sign-reversal in the thermopower around 80 K upon cooling despite the sample having only n-type carriers. On the other hand, the lightly doped sample ($4.9\times10^{19}$ cm$^{-3}$) exhibits only a negative thermopower down to 2 K. These results advance the understanding of Umklapp electron-phonon drag in oxides and highlight KTaO$_3$'s potential for engineering unconventional thermoelectric materials.

## I. INTRODUCTION.

The quest for materials with high thermoelectric efficiency at low temperature has recently resulted in thermoelectric figures of merit exceeding 2.5 [1]. Achieving high thermoelectric performance requires enhancement of the Seebeck coefficient, which can be accomplished through various mechanisms, including the use of 1-dimensional features [1], but also phonon drag. Phonon-drag is an electronic transport phenomenon in which the electron population in a crystal interacts with its phonon population in a manner intense enough that both populations are brought outside thermodynamic equilibrium. This results in a strong enhancement of the Seebeck coefficient over its value in the diffusion regime, where both populations remain in equilibrium. For this to happen, the electron-phonon interactions must be stronger (i.e. have a shorter relaxation time) than the ubiquitous phonon-phonon interactions. Phonon-drag typically results in a boost in thermopower that has a non-monotonic temperature dependence. At temperatures below a maximum that is roughly at the Debye temperature divided by 5, it increases with the increase in phonon population upon warming. At temperatures above that maximum, it decreases as phonon-phonon Umklapp scattering becomes dominant over phonon-electron interactions. In almost all metallic crystals in which conduction is dominated by charge carriers of a single polarity (thus excluding semimetals), the phonon-drag thermopower has the same sign as the diffusion thermopower. The latter is also directly related to the polarity of the charge carrier in most single-polarity metallic solids (except Cu, Ag, Au, and Li, metals in which the Fermi surface is topologically not simply-connected, which gives a positive diffusion thermopower even though the conduction is via electrons) [2] i.e. electrons give a negative diffusion thermopower and holes a positive one. There is only one known case where the phonon-drag thermopower and the diffusion thermopower have opposite signs: Rb [3]. In this work, we report a change in polarity between the diffusion and phonon-drag thermopowers of heavily doped $KTaO_3$ (KTO) films.

The observed change in polarity is attributed to electron-phonon Umklapp scattering. Conventionally, Umklapp (U) scattering is known to play an important role in 3-phonon interaction processes, where it then strongly affects the lattice thermal conductivity. What is much less appreciated is that it also exists for electron-phonon interactions, a fact that is belied by the results presented here. In Normal (N) electron-phonon scattering processes, the change in the electron momentum is equal to that gained by the phonon. When the wave vectors of the interacting electron (**k**) and phonon (**q**) are small, the k-vector of the scattered electron (**k′**) is given by:

$$\mathbf{k'} = \mathbf{k} + \mathbf{q} \qquad (1)$$

and **k′** still falls in the same Brillouin zone (BZ) and its sign is the same as that of **k (Fig. 1(a))**. U-scattering arises when **k** is large enough so that **k** + **q** falls in the next BZ, which happens when |**k** + **q**| > |**g**/2|, where **g** is the reciprocal lattice vector. Now:

$$\mathbf{k'} = \mathbf{k} + \mathbf{q} - \mathbf{g} \qquad (2)$$

where **k′** undergoes a Bragg reflection, leading to a large change of momentum characterized by the vector **g (Fig. 1(b))**. In the diffusive regime, both N and U electron-phonon scattering processes are resistive, and differentiating between them is not critical to either resistivity or thermopower. In case of U-phonon drag processes, the resulting electron (**k′**) will have a momentum opposite to both electrons and phonons due to the Bragg reflection. This manifests itself by generating a Seebeck coefficient of polarity opposite to that of the charge carrier.

While this mechanism was speculated to be true in Rb [4], which has a constant electron density, the present paper offers a more convincing argument, since the doping level of the $KTaO_3$ films can be changed, allowing for a change in **k**. In this work, we present two different Ba-doped $KTaO_3$ with different carrier concentrations, $4.9 \times 10^{19}\ cm^{-3}$ and $3.7 \times 10^{20}\ cm^{-3}$.

There is no change in sign between diffusion and phonon-drag thermopower in lightly doped KTO, where the Fermi surface (**Fig. 1(a)**) is small and |**k** + **q**| < |**g**/2|. In contrast there is one in the heavily doped material, where the condition |**k** + **q**| > |**g**/2| is met (**Fig. 1(b)**).

For completion, we mention the case of $PdCoO_2$ where a positive diffusion thermopower is accompanied by a negative phonon-drag thermopower [5]. Nevertheless, that highly anisotropic material deviates strongly from a simple metal, and the in-plane thermopower is positive while the out-of-plane one is negative [6], making it a goniopolar material [7].

$KTaO_3$ is an incipient ferroelectric that retains a cubic structure ($m\bar{3}m$) down to low temperatures [8, 9]. Stoichiometric $KTaO_3$ is an insulator with a bandgap of ~3.6 eV, separating oxygen 2$p$-derived valence band from tantalum 5$d$-derived conduction band [10]. Due to strong spin-orbit coupling, the six-fold degeneracy of the conduction band's $t_{2g}$ states at the zone center is lifted, resulting in a splitting into total angular momentum states $J = 3/2$ and $J = 1/2$ [11]. In doped KTaO$_3$, itinerant electrons occupy the lower-energy $J = 3/2$ states, which splits into two doubly degenerate bands at finite momentum (depicted as inner and outer bands in **Fig. 1**).

KTaO$_3$ hosts a rich variety of exotic emergent phenomena and properties, including an anisotropic superconducting state that is remarkably resilient to applied magnetic fields [12, 13, 14], non-trivial quantum geometrical effects [15, 16, 17, 18], and efficient spin-to-charge interconversion [19, 20]. These properties are typically probed using magneto-electric experiments. The thermoelectric transport behavior of electron-doped KTaO$_3$ has been explored in bulk materials but not in thin films [21].

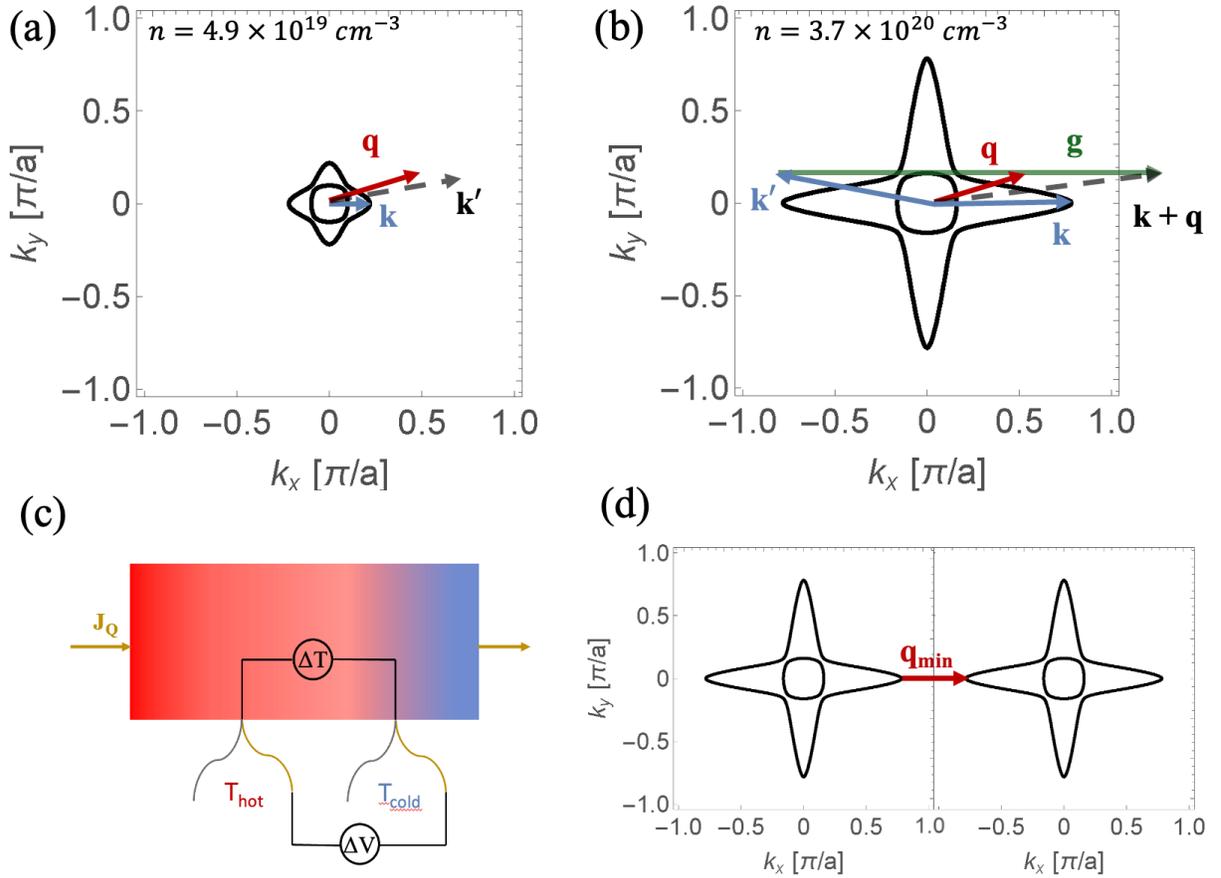

*Figure 1 (a) Normal and (b) Umklapp conditions for electron-phonon scattering. Calculated Fermi surfaces for electron-doped KTaO$_3$ with carrier concentrations, $4.9 \times 10^{19}\ cm^{-3}$ in (a) and*

$3.7 \times 10^{20}$ cm$^{-3}$ *in (b). A "thermal phonon" of energy $k_B T$ and momentum* **q** *interacting with an electron on the Fermi surface of momentum* **k** *does not fulfill the condition for Umklapp scattering in (a) but does in (b). The resulting electron momentum* **k'** *moves in the same direction as* **k** *in (a) but in the opposite direction in (b), where it gives a phonon-drag Seebeck coefficient of polarity opposite to that of the charge carrier. (c) shows a schematic of the sample mount and the direction of the applied temperature gradient. The minimum phonon momentum that satisfies the condition for U-phonon drag is* **q**$_{min}$ *shown in (d).*

## II. METHODS

**Material Growth**

Epitaxial thin films of KTaO$_3$ were grown using a modified Vecco Gen 10 MBE system, where the conventional SiC heater was replaced by 10 μm CO$_2$-laser from Epiray GmbH (THERMALAS Substrate heater). A molecular beam of TaO$_2$ (gas) flux was generated from an effusion cell containing Ta$_2$O$_5$ (H.C. Stark 99.99%) contained in an iridium crucible. TaO$_2$ is the most volatile species in the growth temperature range [22]. Potassium was evaporated from an effusion cell containing an indium-rich (about 4:1 In:K ratio) mixture of potassium and indium so it forms the air-stable intermetallic, In$_4$K. The K-In alloy was prepared in a glove box and contained in a titanium crucible. Once prepared, it can be exposed to air, facilitating its handling and loading. The vapor pressure of potassium is more than 10 orders of magnitude higher than indium at the K-In cell temperature of 300-400 °C [23]. TbScO$_3$ (110)o (Crystec GmbH) and KTaO$_3$ (MTI Corporation) substrates were used as received, where the O subscript indicates orthorhombic indices. Films were grown by co-deposition of potassium, TaO$_2$, and ozone, at substrate temperature of 900 °C as measured by an optical pyrometer operating at a wavelength of 7.5 µm. The films were grown at a background pressure of $1\times10^{-6}$ Torr (10% $O_3$ and 90% $O_2$) and the K:Ta flux ratio was kept at approximately 10:1. A more detailed description is published elsewhere [24]. Fluxes for the sources were $(3-4) \times 10^{13}$ atoms/cm²/s for $TaO_2$ and $(2-3) \times 10^{14}$ atoms/cm²/s for potassium determined by a Quartz crystal microbalance (QCM). Barium fluxes were determined to be $1\times 10^{13}$ atoms/cm²/s at 500 °C and adjusted to doping concentrations with a calibrated vapor pressure curve [24]. An accuracy of ± 15 % can be typically achieved by QCM flux calibration.

The results for the X-Ray Diffraction (XRD) measurements are shown in **Fig. 2 (a).** Also shown are the characterization of the samples by Hall measurements (**Figs. 2 (b) and (c)**). In particular, the samples grown on the $KTaO_3$ (100) substrate had a carrier concentration ($n$) of $3.7\times10^{20}$ cm$^{-3}$ and carrier mobility of 5 cm² V$^{-1}$ s$^{-1}$, while the one grown on $TbScO_3$ (110)$_o$ had carrier concentration around one order of magnitude less ~ $4.9\times10^{19}$ cm$^{-3}$ and carrier mobility of 15 cm² V$^{-1}$ s$^{-1}$.

**Transport Measurements**

Each sample was prepared for transport measurements by attaching the sample to a copper platform atop an alumina base acting as a heat sink. A strain gauge which supplied heat to the sample was adhered to the opposite side of the sample. Two type T thermocouples consisting of a copper and a Constantan wire were placed on one side of the sample to measure the temperature difference. The copper wires were used to measure the thermopower voltage (**Figure 1(c)**). All parts were adhered using silver epoxy. The samples were then placed individually in a vacuumed environment inside a Lakeshore helium-cooled cryostat. Current was supplied to the sample for electrical measurements and strain gauge for thermal measurements via a 6221 Keithley current source. The voltages were read using a 2182A Keithley nanovoltmeter. The residual voltages were less than 2 $\mu$V indicating great electrical contact. The cryostat temperature was controlled via a 331 Lakeshore temperature controller. The temperature was allowed to stabilize for 20 minutes before the sample heater was turned on. The temperature gradient was allowed to stabilize for 5 minutes before thermal measurements.

The value of the uncertainty on the Seebeck coefficient depends on both the noise level for the voltage and the temperature difference between both thermocouples. The uncertainty on the 2182A Keithley nanovoltmeter is 50 nV and with a Seebeck value of 200 µV/K, it is negligible. The uncertainty of the temperature measurement on one thermocouple is 25 mK, so for the difference between two thermocouples, the uncertainty is 50 mK. We apply a temperature difference at room temperature of 3 K, making the uncertainty on the Seebeck coefficient at 300 K around 2 %. At 100 K, the sensitivity of the type T thermocouple is lowered by one half. Additionally, because the sample thermal conductivity increases, and almost reaches its maximum value, the applied temperature gradient is lowered to about 1 K. This results in an uncertainty of about 10 % below 100 K.

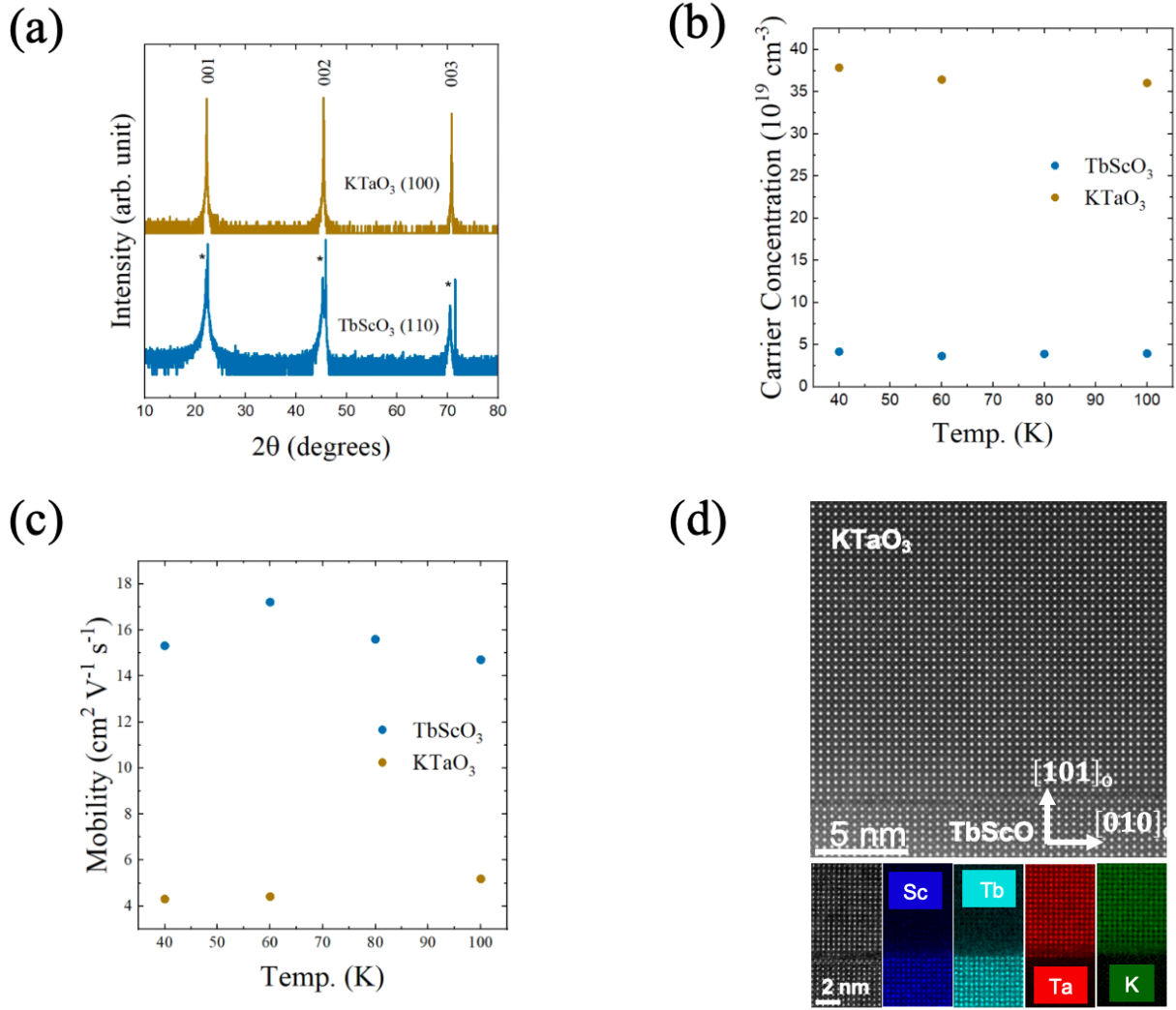

*Figure 2.(a)* shows the XRD measurement for the doped $KTaO_3$ thin film grown on two different substrates, $KTaO_3$ (100) and $TbScO_3$ (110)$_o$. XRD confirms that the samples are single phase and that the $KTaO_3$ films are aligned with the (l00) plane parallel to the surface of the substrate. In the $TbScO_3$ (110)$_o$ sample, we can clearly see the $KTaO_3$ thin film peaks aligning with the $KTaO_3$ substrate sample along with the extra peaks from the $TbScO_3$ (110)$_o$ substrate peaks. *(b)* shows the carrier concentration of the Ba-doped $KTaO_3$ thin film on both substrates at various temperatures. The samples grown on the $KTaO_3$ (100) substrate had n = $3.7 \times 10^{20}$ $cm^{-3}$, while the one grown on $TbScO_3$ (110)$_o$ had around one order of magnitude less n ~ $4.9 \times 10^{19}$ $cm^{-3}$. *(c)* shows the *carrier mobility for both samples. **(d)** shows Scanning Transmission Electron Microscopy (STEM) of the grown KTaO₃ thin film on TbScO₃ (110)ₒ substrate.*

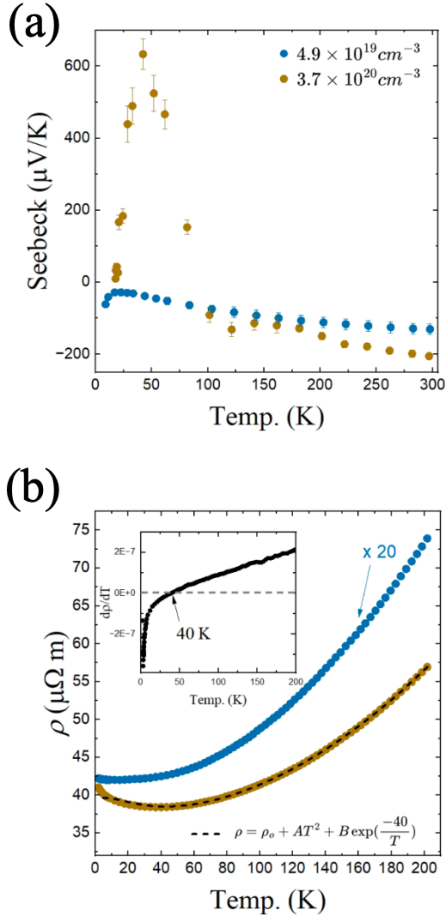

***Figure 3. (a)** shows the thermopower measurements for the doped KTaO₃ thin films with two different carrier concentrations. The sample with lower carrier concentration ($4.9 \times 10^{19}$ $cm^{-3}$) displayed a small negative phonon drag effect at low temperature illustrated by the downturn of signal at the lowest temperatures. The behavior of the thin film with the higher carrier concentration ($3.7 \times 10^{20}$ $cm^{-3}$) is markedly different: the absolute value of the Seebeck signal linearly decreases until 100 K, the diffusive regime. Around 80 K, it changes sign. Then, it sharply increases until a peak occurs around 40 K. Afterwards, it sharply decreases as the temperature*

*drops. Frame **(b)** displays the resistivity of both samples. The change in resistivity with respect to temperature (dρ/dT) for the higher carrier concentration sample ($3.7 \times 10^{20} cm^{-3}$) is shown in the inset. The minimum of resistivity and the peak in Seebeck drag align well around 40 K. The dashed curve through the data for that sample corresponds to an equation for the resistivity in the presence of Umklapp electron phonon scattering.*

## III. RESULTS and DISCUSSION

The Seebeck effect is shown in **Fig. 3 (a)** for both samples. The diffusion thermopower, measured at *T* > 100 K, is negative for both samples, with an absolute value that is nearly linear in *T*, as expected for an *n*-type degenerately-doped semiconductor. The non-monotonic temperature-dependence points to the existence of phonon-drag below 100 K for the sample grown on $KTaO_3$, and below 20 K for the one grown on $TbScO_3$ (110)$_o$. Unexpectedly, the data for the sample on $KTaO_3$ displays a positive thermopower below ~80 K even though there are no holes present in the system. As explained in the introduction, this peculiar observation points to the addition of unexpected scattering events that reverse the momentum direction of the electrons.

The presence of electron-phonon U-phonon drag depends on a few factors: size of the Fermi surface, the mode-and-energy dependent phonon density, and the defects in the system. For phonon drag to occur, the scattering between phonons and electrons of the system needs to be more intense than all other phonon scattering mechanisms, particularly phonon-defect and phonon-phonon scattering. The latter condition implies that phonon-drag occurs at lower temperatures, typically one-fifth of the Debye temperature [4]. Secondly, low defect concentration is needed to minimize phonon-defect scattering and increase the chance of electron-phonon scattering. In the case of our samples, we calculated the Fermi surface for both carrier concentrations using a tight-binding model fit from the ab initio band structure [25]. In both cases, the Fermi surface has two bands, a uniform inner band, and a strongly warped outer band, shown in **Fig. 1**.

As illustrated in **Fig. 1 (d)**, a minimum phonon wave vector (**q$_{min}$**) is required to satisfy the momentum conservation condition for the electron–phonon Umklapp process. To estimate its magnitude, we refer to the phonon band structure of KTaO$_3$ [26], focusing on the longitudinal acoustic (LA) and transverse acoustic (TA) phonon modes. The LA mode reaches a maximum energy of 4 THz (~ 180 K), while the TA mode's maximum energy is at 2 THz (~ 90 K). For the higher carrier concentration sample, the minimum phonon wave vector required to meet the Umklapp condition (|**k** + **q**| > |**g/2**|), is estimated to be ~ 0.2 of the Brillouin zone (BZ). The corresponding phonon energies at this q-value are ~ 80 K for LA and ~ 40 K for TA modes. Both modes contribute to U-phonon drag, but the TA mode is more populated at any temperature and U-phonon-phonon interactions limit the phonon-drag more at 80 K than at 40 K. Therefore, the TA mode is expected to contribute most and the U-phonon drag contribution to the Seebeck coefficient is expected to have a maximum near 40 K, as observed (**Fig. 3 (a)**). Below this temperature, the number of phonons with that energy drops exponentially, causing the sharp decrease in Seebeck.

In contrast, the low carrier concentration sample requires a much larger minimum phonon wave vector (~ 0.8 BZ) to achieve Umklapp scattering. At this q-value, the phonon energies approach the Debye temperatures of the respective modes—approximately 180 K for LA and 90 K for TA. At these higher energies, phonon–phonon U-scattering [27] overwhelms electron-phonon scattering.

Further, phonon-electron U-processes are resistive since the electrons undergo a Bragg reflection. U-phonon drag is thus expected to induce an increase in electrical resistivity [28]. Such a change is clearly observed in **Fig. 2 (b)**. At $T < 40$ K, where the Seebeck coefficient peaks, in the higher carrier sample, resistivity decreases with increasing temperature. This is contradictory to the behavior of normal metals and degenerately doped semiconductors, where resistivity increases with temperature, which is observed at $T > 40$ K. To further investigate this, we fit the data with the following function [28]:

$$\rho = \rho_0 + AT^2 + B\exp\left(\frac{-\Theta}{T}\right) \quad (3)$$

The second term of the equation ($AT^2$) represents the equilibrium electron-electron and electron-phonon scattering [29], while the exponential term represents the U-phonon-drag scattering. Θ is the minimum temperature required for the phonon mode to satisfy the Umklapp scattering condition, i.e. the mode with $q_{min}$ in **Fig. 1 (d)** and an energy of Θ = 40 K as we previously determined. The addition of the Umklapp term in eq. 3 gives the dashed curve in **Fig. 3 (b)** and explains very well the peculiar rise in resistivity below 40 K. The effect is absent in the low-doped sample. This is further evidence for electron-phonon Umklapp scattering in the highly doped sample.

Another detail to consider is the fact that Sakai et al. did not observe any phonon-drag or sign-change in their thermopower measurements on Ba-doped $KTaO_3$ [21]. These were bulk samples grown via the self-flux method and had carrier concentrations from mid-$10^{18}$ to low $10^{20}$ $cm^{-3}$. Additionally, electrical resistivity of all their samples exhibited normal electron scattering that fitted a $T^2$ behavior ($\rho = \rho_0 + AT^2$). Our thin films had higher carrier concentrations than the bulk samples in the aforementioned study [21], and therefore a larger Fermi surface, enabling Umklapp scattering. Additionally, no phonon drag was observed in all of the bulk grown samples, indicating weak electron-phonon interaction [21].

A previous investigation into phonon drag in thin films demonstrated that substrate acoustic phonons can propagate from substrate into the thin films where they interact with the charge carriers, thereby generating phonon drag [30]. The study reported that the phonon drag contribution is generally less pronounced in bulk samples compared to thin films, which is consistent with our observations. Furthermore, it showed that phonon drag signal is influenced by the Debye temperature of the substrate. In our study, both $KTaO_3$ and $TbScO_3$ substrates have comparable Debye temperatures [27]; hence, the interfaces are expected to be quite transparent to the acoustic phonons.

The above discussion is concerned solely with the acoustic phonon modes as they typically have higher sound velocity and higher density of states compared to the optical modes. In the case of $KTaO_3$, the lowest transverse optical mode ($TO_1$) exhibits a peculiar behavior; it has a relatively high sound velocity, comparable to that of the acoustic phonon mode, enabling it to have a potential contribution to the thermal transport processes [26] at very high temperature. The $TO_1$ mode undergoes significant softening at the $\Gamma$ point with decreasing temperature [31]. At 100 K, the mode's energy is 78 K, which lies below the thermal energy of the system and is still populated, but exponentially less so than the TA phonons. As temperature decreases to 40 K, the $TO_1$ mode softens further to energy of 42 K. Even though the energy of the $TO_1$ mode at $\Gamma$ is accessible in the range where thermopower changes sign, no softening has been confirmed at X, making the energy at higher *k*-points not accessible at low temperature. For Umklapp scattering, high *k*-vectors play a significant role. Therefore, acoustic phonons are more likely to be the main players in this scattering event.

## V. CONCLUSIONS

We observed the reversal of phonon-drag thermopower polarity in Ba-doped $KTaO_3$ thin films mediated by electron-phonon Umklapp scattering. The polarity reversal occurs in a heavily doped sample (carrier concentration of $3.7 \times 10^{20}$ cm$^{-3}$) where the Fermi surface spans 0.8 of the Brillouin zone, enabling electron-phonon Umklapp scattering to occur which reverses the electron momentum. This manifests as a positive thermopower below 80 K despite the *n*-type conduction, contrasting with conventional behavior in the lightly doped sample ($4.9 \times 10^{19}$ cm$^{-3}$) that maintained negative thermopower down to 2 K. The successful epitaxial growth of $KTaO_3$ films enables precise doping control, providing a platform to explore electron-phonon phenomena in quantum materials.

## Acknowledgement

Primary support for this work (MN, KA and JPH) comes from the Center for Emergent Materials, a National Science Foundation (NSF) MRSEC, grant number DMR 2011876. BLW was supported by the DoD SMART scholarship. SJP was supported by the U.S. NSF under Grant DMR-2408890. This work made use of the synthesis and electron microscopy facilities of the Platform for the Accelerated Realization, Analysis, and Discovery of Interface Materials (PARADIM), which are supported by the NSF, Cooperative Agreement No. DMR-2039380. M.N.G is supported by the Ramon y Cajal Grant RYC2021-031639-I funded by MCIN/AEI/ 10.13039/501100011033.

## Author contributions

B.L.W., K.A., and J.P.H. conceived the project; M.N., S.P. and B.L.W. performed thermoelectric and electric measurements, and analyzed the data with help from J.P.H and K.A.; S.P. and T.S. grew and characterized the samples with help from K.A. and D.G.S.; M.N.G. calculated the Fermi surface of the samples; S.S.R. conducted TEM; M.N., B.L.W., K.A. and J.P.H. wrote the manuscript with input from all coauthors.

## Competing interests

The authors declare no competing interests.

## Data Availability

Data are available from the corresponding authors upon reasonable request.

## Corresponding authors

Correspondence to Brandi L. Wooten and/or Joseph P. Heremans.